\documentclass[12pt]{article}
\begin{document}
\pagenumbering{arabic}
\title{Gravitational energy of a magnetized Schwarzschild black hole - a teleparallel approach}

\author{K. H. C. Castello-Branco$\,^{1\ast}$ \\ and
J. F. da Rocha-Neto$\,^{2\dagger}$}
\date{}
\maketitle

\begin{center}
1 Universidade Federal do Oeste do Par\'a, \\
Av. Marechal Rondon, 68040-070. Santar\'em, PA, Brazil.
\end{center}

\begin{center}
2 Instituto de F\'isica, Universidade de Bras\'ilia, \\
70910-900, Bras\'ilia, DF, Brazil.
\end{center}

\begin{abstract}
We investigate the distribution of gravitational energy in the spacetime of a 
Schwarzschild black hole immersed in a cosmic magnetic field. This is done in the 
context of the {\it Teleparallel Equivalent of General Relativity}, which is an alternative geometrical 
formulation of General Relativity, where gravity is described by a spacetime endowed with torsion rather than curvature, 
whose fundamental field variables are tetrad fields. We calculate the 
energy enclosed by a two-surface of constant radius - in particular, the energy enclosed by the event horizon of the black hole. 
In this case we find that the magnetic field has the effect of increasing 
the gravitational energy as compared to the vacuum Schwarzschild case. We also compute the energy (i) in the weak magnetic field limit, 
(ii) in the limit of vanishing magnetic field, and (iii) in the absence of the black hole. In all cases our results are consistent 
with what should be expected on physical grounds.  
\end{abstract}

Keywords: {\it teleparallelism; tetrad field; gravitational energy; magnetized Schwarzschild black hole}

PACS numbers: 04.20.-q, 04.20.Cv

{\footnotesize
\noindent $\ast$ khccb@yahoo.com.br\\
\noindent $\dagger$ rocha@fis.unb.br}

\bigskip
\section{Introduction}
Black hole physics has reached an impressive stage of development, even though 
a complete understanding of it is expected to be achieved only by means of a 
full theory of quantum gravity. Although the general main properties of black holes
were obtained through investigations basically concerning isolated and asymptotically 
flat black holes, it is certainly important to understand the interaction of a black 
hole with the astrophysical environment that surrounds it. In fact, models of 
astrophysical black holes are needed in order to describe, for instance, 
(i) the effect on the black hole near-horizon region of an external magnetic field 
produced by currents in an accretion disk or (ii) the mechanism 
responsible for the huge amount of energy released in active galactic nuclei 
containing a supermassive black hole. In the first case, as an approximation, 
one can model that effect by simply considering a Schwarzschild black hole 
immersed in an external magnetic field. The corresponding metric is described 
by the Schwarzschild-Melvin black hole solution \cite{Ernst}, which is the unique exact, 
static, axisymmetric black hole solution of the sourceless Einstein-Maxwell equations 
that asymptotically approaches Melvin's magnetic universe (MMU) \cite{Hiscock}.

The MMU is a solution of the Einstein's equation that describes a matter-free universe,
 endowed only with a magnetic field \cite{BM1},\cite{BM2},\cite{Melvin1},\cite{Melvin2}. 
Of course, the Schwarzschild-Melvin black hole solution is not asymptotically flat, with the 
asymptotic cosmological magnetic field strength appearing in the solution together with the black hole 
mass as one of the two characteristic parameters of the corresponding metric. This solution can provide 
insights in understanding more realistic situations and this has just been the motivation of several 
current studies in the literature about the Schwarzschild-Melvin solution as well 
as other Melvin-like black hole solutions. As remarked by, {\it e.g.}, Konoplya \cite{konoplya}, 
it is known that the large-scale magnetic field in Universe
has poloidal and toroidal components and the dominant component is the poloidal one, what implies that 
the Schwarzschild-Melvin solution, with the magnetic field specifying a single direction in space, can be considered 
as a reasonable approximation. Besides, in what concerns astrophysical motivations, it has been addresed 
that the Schwarzschild-Melvin solution is valid for estimations about lens effects 
due to the magnetic field of that kind of magnetized black hole \cite{konoplya}. Furthermore, what makes the magnetized 
Schwarzschild solution specially attractive is that it may be the unique exact solution for a magnetized black hole that still 
has a regular event horizon (i.e., the one for which the magnetic field does not spoil the regularity of the black hole 
horizon) \cite{Ernst},\cite{Hiscock}. Also, bearing in mind that isolated, asymptotically flat 
black holes are idealized, in view of astrophysical motivations it is of interest to study a black hole in a magnetic field.

In particular, it is interesting to investigate the gravitational energy of a magnetized black hole. How, for 
instance, the magnetic field affects the distribution of gravitational energy in the near-horizon region of the 
black hole ? In this paper we consider the Schwarzschild-Melvin black hole solution for addressing  
the distribution of gravitational energy associated to it. 
This question has been dicussed in General Relativity by means of pseudotensors 
\cite{Xulu1} and conserved charges \cite{Dadhich}. Here, we consider it in the context of the 
\textit{Telepara-\linebreak llel Equivalent of General Relativity}
 (TEGR) \cite{Hehl},\cite{HM},\cite{Rocha1},\cite{Obukhov1},\cite{Obukhov2},\cite{ChenNester},\cite{Blago},\linebreak 
 \cite{Ortin}, 
which is an alternative geometrical formulation of General Relativity, where gravity is described by a spacetime 
manifold endowed with torsion, with null curvature tensor (Weitzenb\" ock spacetime). In particular, 
we investigate the energy enclosed by the event horizon, an analysis that has been not addressed in Ref.'s 
\cite{Xulu1},\cite{Dadhich}.
Recently, the TEGR  has shown to imply nonlocal modifications of General Relativity 
which might be responsible for the effect attributed to dark matter \cite{Hehl1}. 
Earlier, it has been shown that the TEGR is a suitable framework for addressing the issue of defining both the 
energy-momentum and angular momentum densities of the gravitational field.
These definitions were first shown to arise from the Hamiltonian formulation of 
the theory \cite{Rocha1}, but later it was shown that the same definition 
for the gravitational energy-momentum vector then established can also be derived directly from the 
field equation of the TEGR \cite{Maluf2}, which turns out to be equivalent to Einstein's field equation 
of General Relativity. Recently, that definition has 
been applied to the standard cosmological spacetimes \cite{Ulhoa}. 
The basic field variables of the TEGR are the {\it tetrad} fields (rather than the metric), which can 
naturally be interpreted as {\it reference frames} adapted to observers in spacetime \cite{Synge}. 
This interpretation has been explored in investigations on both the energy 
and angular momentum of the gravitational field in several important spacetimes \cite{MalufCQG},\cite{Maluf-Ulhoa-PRD2009}. 

This paper is organized as follows. In Sec. II we review the definition for
the gravitational energy-momentum established in the context of the 
TEGR; in Sec. III we review the magnetized Schwarzschild black hole solution; 
in Sec. IV we compute the energy enclosed by a two-surface of constant radius and then 
investigate the distribution of gravitational energy on the spacetime of the magnetized Schwarzschild black hole; 
in Sec. V we make the final discussion and remarks.

Throughout the paper the following notation is used: spacetime indices $\mu, \nu, ...$ and SO(3,1) indices $a,
b, ...$ run from 0 to 3. Time and space indices are indicated
according to $\mu=0,i$ and $a=(0),(i)$. The tetrad field is denoted by
$e^a\,_\mu$, and the flat, Minkowski spacetime metric tensor raises
and lowers tetrad indices and is fixed by $\eta_{ab}=e_{a\mu}
e_{b\nu}g^{\mu\nu}= (-1,+1,+1,+1)$. The determinant of the tetrad
field is denoted by $e=\det(e^a\,_\mu)$. 

\bigskip

\section{Gravitational energy-momentum in TEGR}

\noindent

Basically the TEGR is just a reformulation of Einstein's general relativity in terms of tetrad 
fields, which are the basic variables of theory, and the torsion tensor. The field equation for the tetrad field 
is equivalent to Einstein's field equation of general relativity (see in the following). Therefore, the TEGR is just an 
alternative geometrical formulation of general relativity, rather than a new theory of gravity. Hence, it  
can provide new insights about the behaviour of gravity. In particular, it has been shown that the TEGR is a suitable 
framework for addressing both the energy-momentum and angular momentum densities of the gravitational 
field \cite{MalufCQG},\cite{Maluf-Ulhoa-PRD2009},\cite{Maluf-Ulhoa-angmom}.
 
The equivalence of the TEGR with 
Einstein's general relativity is achieved by means of the following identity between the scalar curvature 
$R(e^a\,_\mu)\,$, constructed out of the tetrad field, and a combination 
of quadratic terms of the torsion tensor, namely (see, {\it e.g.}, \cite{Maluf2})

\begin{equation}
eR(e^a\,_\mu)\equiv -e(\frac{1}{4}T^{abc}T_{abc}+\frac{1}{2}T^{abc}
T_{bac}-T^aT_a)
+2\partial_\mu(eT^\mu)\,.
\label{ident-esc-curvat}
\end{equation}
For reviews on the formulation of Einstein's general relativity in the context of the 
teleparallel geometry we  refer the reader to, {\it e.g.}, \cite{Hehl},\cite{HM},
\cite{Blago},\cite{Ortin},\cite{Maluf2}.

We remark that the teleparallel description of gravity is not unique, since it can either be described 
by a Lagrangian that is invariant under the local \cite{Hehl},\cite{HM},\cite{Obukhov1} or global Lorentz 
($SO(3,1)$) group \cite{Rocha1},\cite{Hay},\cite{Hayashi-Shirafuji1},\cite{Hayashi-Shirafuji2},\cite{Nester2},\cite{Pereira97}. In the 
context of metric-affine theories of gravity, one naturally demands local Lorentz invariance, inspired by 
the crucial role this symmetry has on quantum field theory, with gravity thus been considered as 
a gauge theory of the Poincar\'e group \cite{HM}. Nevertheless, {\it a priori}, there is no physical reason for ruling 
out theories of gravity which exibits invariance under {\it global} $SO(3,1)$ symmetry. In particular, one can consider 
a teleparallel theory of gravity which has this kind of symmetry, in which case one may consider the TEGR as a gauge 
theory for the translation group \cite{Pereira97}. In the 
TEGR with global $SO(3,1)$ symmetry, although two sets of tetrad fields which are related by a local $SO(3,1)$ 
transformation yield the same metric tensor, they are not physically equivalent. In this case, one has to handle with 
the six extra degrees of freedom of each tetrad field. It has been shown that these can naturally be fixed 
by exploring the interpretation of tetrad fields as reference frames adapted to observers in spacetime 
\cite{MalufCQG},\cite{Maluf-Ulhoa-PRD2009}. Two sets of tetrad fields related by a local $SO(3,1)$ transformation 
represent reference frames which have different translational and rotational accelerations. Here we consider the 
TEGR with global $SO(3,1)$ symmetry. In this case the spin ({\it i.e.}, $SO(3,1)$) connection is dropped out, what 
implies that the torsion tensor is simplified  to (see, {\it e.g.}, \cite{Maluf2})
\begin{equation}
T_{a\mu\nu} = {\partial}_{\mu}e_{a\nu}-{\partial}_{\nu}e_{a\mu}\,,
\label{torsion}
\end{equation}
depending now only on the tetrad field.
It should be noted that every tetrad field that is a solution of the theory with local $SO(3,1)$ symmetry is 
also a solution of the theory with global $SO(3,1)$ symmetry, what means that the absence of local $SO(3,1)$ symmetry 
does not imply any restriction on the possible tetrad fields \cite{Maluf2}. 

The Lagrangian density of the TEGR is given by the combination of the 
quadratic terms on the right-hand side of  Eq. (\ref{ident-esc-curvat}), 

\begin{eqnarray}
L&=& -k e(\frac{1}{4}T^{abc}T_{abc}+\frac{1}{2}T^{abc}T_{bac}-
T^aT_a) - \frac{1}{c}L_m\nonumber \\
&\equiv& -ke\Sigma^{abc}T_{abc}-\frac{1}{c}L_m\,, 
\label{lagrang-tegr}
\end{eqnarray}
where $k=c^{3}/16\pi G\,$, $T_a=T^b\,_{ba}\,$, $T_{abc}=e_b\,^\mu e_c\,^\nu T_{a\mu\nu}\,$ and $\Sigma^{abc}$ is defined 
by

\begin{equation}
\Sigma^{abc}= \frac{1}{4} (T^{abc}+T^{bac}-T^{cab})
+\frac{1}{2}( \eta^{ac}T^b-\eta^{ab}T^c)\,.
\label{def-sigma}
\end{equation}
$L_m$  is the Lagrangian density for matter fields. 
 
The field equation for the tetrad field derived from (\ref{lagrang-tegr}) is equivalent to Einstein's 
equation, and it reads

\begin{equation}
e_{a\lambda}e_{b\mu}\partial_\nu (e\Sigma^{b\lambda \nu} )-
e (\Sigma^{b\nu}\,_aT_{b\nu\mu}-
\frac{1}{4}e_{a\mu}T_{bcd}\Sigma^{bcd} )=\frac{1}{4kc}eT_{a\mu}\,,
\label{eq-campo-tegr}
\end{equation}
where
$eT_{a\mu}=\delta L_m / \delta e^{a\mu}$. 
In fact, it is possible to show that the left-hand side of Eq. (\ref{eq-campo-tegr}) 
may be rewritten exactly as $\frac{1}{2}e\left[ R_{a\mu}(e)-\frac{1}{2}e_{a\mu}R(e)\right]$. Thus it turns out  
that (\ref{eq-campo-tegr}) is the Einstein's equation of general relativity in terms of tetrad fields. From now on we 
will set $c=G=1$, unless otherwise these constants are explicitly shown.

As shown in Ref. \cite{Maluf2}, Eq. (\ref{eq-campo-tegr}) may be simplified as 

\begin{equation}
\partial_\nu(e\Sigma^{a\lambda\nu})=\frac{1}{4k}
e\, e^a\,_\mu( t^{\lambda \mu} + T^{\lambda \mu})\;,
\label{5}
\end{equation}
where $T^{\lambda\mu}=e_a\,^{\lambda}T^{a\mu}$ and
$t^{\lambda\mu}$ is defined by

\begin{equation}
t^{\lambda \mu}=k(4\Sigma^{bc\lambda}T_{bc}\,^\mu-
g^{\lambda \mu}\Sigma^{bcd}T_{bcd})\,,
\label{6}
\end{equation}
and in view of the property 
$\Sigma^{a\mu\nu}=-\Sigma^{a\nu\mu}$ it follows that

\begin{equation}
\partial_\lambda
\left[e\, e^a\,_\mu( t^{\lambda \mu} + T^{\lambda \mu})\right]=0\,.
\label{7}
\end{equation}
This equation then yields the following continuity (or balance) equation,

\begin{equation}
\frac{d}{dt} \int_V d^3x\,e\,e^a\,_\mu (t^{0\mu} +T^{0\mu})
=-\oint_S dS_j\,
\left[e\,e^a\,_\mu (t^{j\mu} +T^{j\mu})\right]\,.
\label{8}
\end{equation}
Thus $t^{\lambda\mu}$ can be identified as the {\it gravitational energy-momentum tensor} 
\cite{Maluf2},\cite{Maluf-Faria-CB2003} and

\begin{equation}
P^a=\int_V d^3x\,e\,e^a\,_\mu (t^{0\mu} 
+T^{0\mu})
\label{9}
\end{equation}
as the total energy-momentum contained within a volume $V$ of the three-dimensional space.
In view of (\ref{5}), Eq. (\ref{9}) may be written simply as 

\begin{equation}
P^a=-\int_V d^3x \partial_i \Pi^{ai}\quad,
\label{def-energia-mom}
\end{equation}
where $\Pi^{ai}=-4ke\,\Sigma^{a0i}$. $\Pi^{ai}$ is the momentum
canonically conjugated to $e_{ai}$ (see \cite{Maluf5}). We remark that expression (\ref{def-energia-mom}) is exactly the definition for the gravitational energy-momentum presented in Ref. \cite{Rocha1}, established in the context of the Hamiltonian formulation of the TEGR, by taking 
the integral form of the constraint equations. However, we have reviewed here how it is obtained directly from the field equation 
of the TEGR (see \cite{Maluf2} and \cite{Maluf-Faria-CB2003}), independently of the Hamiltonian formulation. Anyway, the Hamiltonian formulation 
was decisive in identifying the form of $P^a$ \cite{Rocha1}. 

If one considers the $a=(0)$ component of Eq. (\ref{def-energia-mom}) and adopts asymptotic 
boundary conditions for the tetrad field it results \cite{Rocha1} that the ensuing expression is precisely the surface integral at infinity that defines the ADM energy. Also, when applied to the calculation of the gravitational energy of the Kerr black hole such expression has led to very close agreement with the irreducible mass of the black hole, wihout the need to resort to the slow rotation approximation \cite{Rocha1}. These facts, as well as the conclusions that has been addressed, {\it e.g.}, in  \cite{MalufCQG},\cite{Maluf-Ulhoa-PRD2009} are a strong indication that Eq. (\ref{def-energia-mom}) does indeed is a viable definition to represent the gravitational energy-momentum. For instance, in Ref. \cite{MalufCQG}, by considering the interpretation of tetrad fields as reference frames in spacetime, it has been shown that for the set of tetrad fields adapted to observers in free fall in the Schwarzschild spacetime the gravitational 
energy-momentum constructed out of this set of tetrad fields vanishes, in agreement with the Equivalence Principle. It follows that the gravitational energy enclosed by a three-dimensional volume, limited by a surface $S$, is  given by the $a=(0)$ component of Eq. (\ref{def-energia-mom}), {\it i.e.},
\begin{equation}
P^{(0)}=  \oint_{S}\, dS_{i}\,4k e\Sigma^{(0)0i}\quad.
\label{energia-grav}
\end{equation}

It must be noted that the evaluation of Eq. (\ref{def-energia-mom}) is carried out in the 
configuration space and that the definition of energy as given by this equation transforms 
as the $(0)$ component of $P^{a}$, which is invariant under general coordinate transformations of the three-dimensional space, under time 
reparametrizations, and under global SO(3,1) transformations as well. The
non-invariance of Eq. (\ref{def-energia-mom}) under the local SO(3,1) group is related to the 
frame dependence of the definition.  As argued in Ref. \cite{MalufCQG}, this dependence 
is a natural feature of $P^a\,$, since in the TEGR any global tetrad frame can be choosen for the 
description of a solution of the Einstein's field equation and each such frame yields a viable 
teleparallel description of the spacetime geometry \cite{Nester}.

\section{Magnetized Schwarzschild Black Hole}

\noindent

Families of magnetized black hole solutions were found by Ernst \cite{Ernst} basically by means of a   
special (Harrison-like) transformation applied to Minkow-\linebreak ski spacetime and then applying the same 
procedure to black hole solutions (for a general account on magnetized black hole solutions, 
see \cite{Stephani}). The metric describing the Schwarzschild black hole in an external 
magnetic field (or simply the magnetized Schwarzschild black hole) can be written in spherical-like 
coordinates as \cite{Ernst},\cite{Hiscock} (see section 22.2 of Ref. \cite{Stephani}, for stationary, 
axially symmetric magnetic solutions)

\begin{equation}
ds^{2} = {\Lambda}^{2}\left [-\left (1-\frac{2m}{r}\right )dt^{2} 
+ \left (1-\frac{2m}{r}\right )^{-1}dr^{2}+r^{2}d\theta^{2}\right ] 
+ {\Lambda}^{-2}r^{2}\sin ^{2}\theta d\phi^{2}\,\,,
\label{3.10} 
\end{equation}
where $\Lambda=1+ \frac{1}{4}B_{0}^{2}r^{2}\sin^{2}\theta\,$. $B_0$ is 
a constant that corresponds to the value of the magnetic field everywhere on the polar axis and 
represents the strength of the cosmological magnetic field. For 
$m=0$, Eq. (\ref{3.10}) yields the metric of the MMU in spherical-like coordinates 
and for $B_{0} = 0$ it reduces to the Schwarzschild metric. 
Although it is not asymptotically flat, but rather resembles MMU, this metric is static  
and thus it can easily be seen that its causal structure is the same as that for the Schwarzschild vacuum 
case ($B_{0}=0$). Thus, the metric (\ref{3.10}) describes a black hole in terms of the usual 
black hole definition. The same kind of identification is not straightforward for other magnetized 
solutions whose spacetimes, as Kerr-Newman-like,  are only stationary. 
Here we will restrict our attention only to the magnetized Schwarzschild solution. 

As the metric of interest here concerns a black hole in a magnetic field, what might be important 
in modelling astrophysical black holes, it is important to have a notion of the characteristic 
scales of the strength of such a magnetic field for those kinds of black holes. Following Frolov 
and Shoom \cite{frolov-shoom}, one can consider estimates given in \cite{piotrovich} for the magnetic 
field near the horizon of a stellar mass ($\sim 10M_{\odot}$) black hole, which is of the order of $\sim 10^{8}\,G$, 
while near the horizon of a supermassive black hole 
($\sim 10^{9}M_{\odot}$) it is of order $\sim 10^{4}\,G$. However, both of these orders of magnitude 
are weak, if compared to the characteristic scale of the strong magnetic field that can distorts the near-horizon 
geometry of a black hole. Indeed, as the local curvature created by a magnetic field $B$ is of order of 
$GB^{2}/c^{4}$, then it is comparable to the spacetime curvature near a black hole of mass $M$ if 
$(GB^{2}/c^{4})\sim 1/(GM/c^{2})^{2}$. Hence, for a black hole of mass $M$ this holds if 
$B\sim B_{M}=\frac{c^{4}}{G^{3/2}M_{\odot}}\left(M_{\odot}/{M}\right)\sim 10^{19}(M_{\odot}/{M})\,G$ 
(see, {\it e.g.}, \cite{frolov-shoom}). We thus see that $B_M$ is much greater than 
the estimated values for the stellar mass and supermassive black holes mentioned above. In this case, one 
usually considers the magnetic field as a test-field in the given black hole background (as in, {\it e.g.}, 
\cite{konoplya} and \cite{frolov-shoom}). Nevertheless, if there exist supermassive black holes with masses 
$\sim 10^{14}M_{\odot}$, then the corresponding magnetic field would be $\sim 10^{5}\,G$, what is a reasonable 
value that might occur in nature.

\section{Energy Distribution}

\noindent

A tetrad field is defined by a set $e^{a}\,_{\mu}$ of four orthonormal, linearly independent vectors in spacetime.  
To each observer in spacetime one can adapt a tetrad field in the following way \cite{Synge}. If $x^{\mu}(s)$ 
denotes the world line $C$ of an observer in spacetime, where $s$ is the observer's proper time, the 
observer's four-velocity along $C$, defined by $u^{\mu}(s)=dx^{\mu}/ds$, is identified with the $a=(0)$ component 
of $e_{a}\,^{\mu}$, that is, $u^{\mu}(s)=e_{(0)}\,^{\mu}$ along $C$. In this way, each set of tetrad fields defines 
a class of referance frames in spacetime \cite{Synge}. 
In what follows we will consider a set of tetrad fields adapted 
to a static observer in spacetime \cite{MalufCQG}. Given a metric $g_{\mu\nu}$, 
the tetrad field related to it can be easily obtained through 
$g_{\mu\nu}=\eta^{ab}e_{b\mu}e_{a\nu}$. The consideration of static 
observers is achieved by imposing on $e_{a\mu}$ 
the conditions (i) $e_{(0)}\,^{i}=0\,$, which implies that $e_{(k)0}=0\,$, and (ii) $e_{(0){i}}=0\,$, 
which implies that $e_{(k)}\,^{0}=0\,$. While 
the physical meaning of condition (i) is straightforward (the translational velocity of the observer is null, \textit{i.e.}, 
the three components of the frame velocity in the 
three-dimensional space are null), for condition (ii) it is not so. The latter is a condition on the 
rotational state of motion of the observer. It implies that the observer (\textit{i.e.}, the
three spatial axes of the observer's local spatial frame) is (are) not rotating 
with respect to a nonrotating frame (for a detailed dicussion, see \cite{MalufCQG} and references therein). Hence, 
conditions (i) and (ii) are six conditions which completely fix the structure of a tetrad field. 

The tetrad field related to metric (\ref{3.10}), and that satisfies both conditions (i) and (ii) discussed above, 
is thus given by

\begin{equation}
e_{a\mu} = \left(\begin{array}{cccc}
-A & 0 & 0 & 0\\
0 & B\sin\theta\cos\phi & \Lambda r\cos\theta\cos\phi & -{\Lambda}^{-1}r\sin\theta\sin\phi \\
0 & B\sin\theta\sin\phi & \Lambda r\cos\theta\sin\phi & {\Lambda}^{-1}r\sin\theta\cos\phi \\
0 & B\cos\theta & -\Lambda r\sin\theta & 0
\label{3.11}
\end{array}
\right),
\end{equation}

where

\begin{eqnarray}
A &= & \Lambda\left (1-\frac{2m}{r}\right )^{1/2},\nonumber\\
B &= & \Lambda\left (1-\frac{2m}{r}\right )^{-1/2}.
\label{3.12}
\end{eqnarray}
It follows that $e=\det(e^a\,_\mu)$ is given by $e = \Lambda^{2}r^{2}\sin\theta$.

From (\ref{energia-grav}), the energy enclosed by a surface of fixed radius $r=r_0$ is given by
\begin{equation}
P^{(0)} = 4 k \int_{S} d\theta d\phi\,e\Sigma^{(0)01}\,.
\label{4.6}
\end{equation}

In order to evaluate $\Sigma^{(0)01}\,$, associatad with the tetrad field given by Eq. 
(\ref{3.11}), we resort to Eq. (\ref{def-sigma}). After algebraic manipulations it yields

\begin{equation}
\Sigma^{(0)01}=A\frac{1}{2}g^{00}g^{11}(g^{22}T_{212}+g^{33}T_{313})\,.
\label{4.3}
\end{equation}
The components of the torsion tensor appearing in the above
expression can be directly calculated and are given by
\begin{eqnarray}
T_{212}&=&-\Lambda^{2} r\left(1 - \frac{2m}{r}\right)^{-1/2} + 
\Lambda r\frac{\partial (r\Lambda)}{\partial r} \,,\nonumber\\
T_{313}&=&-r\sin^2\theta  \left[\left(1 - \frac{2m}{r}\right)^{-1/2} - 
\Lambda^{-1}\frac{\partial(r\Lambda^{-1})}{\partial r}\right]\,.\label{4.4}
\end{eqnarray}
Taking these into Eq. (\ref{4.3}) we obtain
\begin{equation}
e\Sigma^{(0)01}=\Lambda^{-1}r\sin\theta\,\left[\frac{1}{2}
-\left(1-\frac{2m}{r}\right)^{1/2}\right]+\frac{1}{2}\Lambda r\sin\theta\,,
\label{4.5}
 \end{equation} 
where we have made use of the expression $e = \Lambda^{2}r^{2}\sin \theta$.

Thus, after perfoming the integration in Eq. (\ref{4.6}), we are left with
\begin{eqnarray}
P^{(0)}&=&\left(\frac{c^4}{G}\right)\frac{ r_{0}}{4}\biggl\{2 + \left(\frac{G}{c^4}\right)\frac{(B_{0}r_{0})^2}{3} 
+{2\left(1 - 2\sqrt{1 - 2(G/c^2)m/r_0}\right)\over{(\sqrt{G}/c^{2})B_{0}r_{0}\sqrt{\frac{G}{c^4}(B_{0}r_{0})^{2}+4}}} \nonumber\\ 
&\times &\ln\left({{\sqrt{\frac{G}{c^4}(B_{0}r_{0})^{2}
+ 4}+ (\sqrt{G}/c^{2})B_{0}r_{0}}\over{\sqrt{\frac{G}{c^4}(B_{0}r_{0})^{2} + 4} 
- (\sqrt{G}/c^{2})B_{0}r_{0}}}\right)^{2}\biggr\}\,,
\label{4.7}
\end{eqnarray} 
where we have restored the physical constants $G$ and $c$ (note that $r_0$ has dimensions of $Gm/c^{2}$ and $B_0$ of 
$c^{4}G^{-3/2}m^{-1}$). This gives the gravitational energy, due to contributions of the mass and the cosmic magnetic field, 
within a spherical surface of radius $r_{0}$ enclosing the  event horizon. Let us discuss its 
meaning in the following, where we will again set $G=c=1$, for the sake of simplcicity.
 
We first examine the gravitational effect of the magnetic field. In the limit of weak magnetic 
field, that is, if $B_0$ is such that $|B_{0}m|\,\ll 1$, in the region $2m\ll r \ll B_{0}^{-1}\,$ 
we have $\Lambda\approx 1\,$. Thus, in this limit the spacetime outer of the 
horizon is approximately flat and the magnetic field in this region behaves like 
an asymptotically uniform test field on a Schwarzschild background \cite{Ernst}. 
From (\ref{4.7}) it follows that in this approximation we have 
\begin{equation}
P^{(0)}\approx m + \frac{1}{6} B_{0}^{2}r_{0}^{3}\,.
\label{campo-fraco}
\end{equation}
This result corresponds to what should be expected on physical grounds for a black hole immersed in a weak magnetic field. In this case 
the black hole spacetime is not distorted by the magnetic field and thus the energy enclosed by a large surface of 
constant radius should give the total energy of the spacetime, {\it i.e.}, the black hole rest-mass (energy), 
given by the term $m$, plus the magnetic energy enclosed by a surface of radius $r_0$, which is given by 
$B_{0}^{2}r_{0}^{3}/6\,$. We note that approximation (\ref{campo-fraco}) for the energy gives the result 
obtained by Xulu \cite{Xulu1} by means of the Einstein's pseudo-tensor of energy-momentum. This agreement might be 
expected, since pseudo-tensors give the total energy of the spacetime, for the integral that defines 
them is evaluated at a surface of constant radius at large distances from the black hole. 
In the absence of the magnetic field ({\it i.e.}, for $B_{0} =0$), we recover, as expected, the energy value for the 
Schwarzschild vacuum solution, that is, the ADM energy, $m$. Nevertheless, we note that we recover this not only as a 
result of considering the approximation  $2m\ll r \ll B_{0}^{-1}\,$, but in fact also directly from Eq. (\ref{4.7}), 
without any approximation, when we simply take the limit 
$B_{0} =0$ (see Eq. (\ref{4.8}) in the following). 
 
In the general rather than weak magnetic field case, according to Eq. (\ref{4.7}), the effect 
of the magnetic field is stronger. In fact, at the event horizon ($r_{0}=2m$) it gives
\begin{equation}
P^{(0)}= m + \frac{2}{3}B_{0}^{2}m^{3}+\frac{1}{B_{0}\sqrt{B_{0}^{2}m^{2}+1}}\,\ln\left(\frac{\sqrt{B_{0}^{2}m^{2}+1}+ B_{0}m}{\sqrt{B_{0}^{2}m^{2}+1} 
- B_{0}m}\right)\,,
\label{energy-horizon}
\end{equation} 
what shows that the effect of the magnetic field is to increase the energy enclosed by the event horizon, 
as compared to the vacuum Schwarzschild case. By expanding the latter expression for small 
$B_0$ we can see this more clearly, as 
\begin{equation}
P^{(0)} = 2m +(2/3)B_{0}^{2}m^{3} + (8/15)B_{0}^{4}m^{5} + O(B_0^{5})\,.
\label{energy-horizon2}
\end{equation}
For $B_{0}=0\,$, we obtain the energy enclosed by the horizon of the vacuum Schwarzschild 
black hole. This can also be recovered generically, when we take the magnetic field strength $B_{0}$ equal to zero in 
Eq. (\ref{4.7}), which is an exact result. In this case, it reduces to
\begin{equation}
P^{(0)} = r_{0}\left(1 - \sqrt{1 - 2m/r_{0}}\right)\,,\label{4.8}
\end{equation}
which is the energy contained in a spherical surface of radius $r_{0}$ in the Schwarzschild spacetime. 
This result is the same obtained  by Maluf \cite{Maluf95} for the Schwarzschild solution, but in the 
context of the Hamiltonian formulation of the TEGR. However, here it is obtained as a particular 
case of the magnetized Schwarzschild black hole solution and, most important, from a definition 
for gravitational  energy-momentum that arises solely from the field equations of the TEGR, 
{\it independently of its Hamiltonian formulation}. We also note that result of Eq. (\ref{4.8}) 
is the same achieved by Brown and York \cite{BY}, in the context of their quasilocal energy 
analysis. In particular, if we now take in Eq.  (\ref{4.8}) $r_{0} = 2m\,$, we have the energy 
inside of the event horizon of the Schwarzschild black hole, that is, $P^{(0)} = 2m\,$.

Finally, if we take the limit of $r_{0}$ going to infinity in Eq. 
(\ref{4.7}), the logarithmic term goes to zero. However,  because of the first and 
second terms, $P^{(0)}$ will be infinite.  This result reflects the 
fact that in the solution of the magnetized Schwarzschild black hole there exists a cosmological  
magnetic field, what corresponds to an infinite amount 
of energy. Thus in the region very far away from the black hole the energy that 
predominates is the magnetic energy. In fact, as $r\longrightarrow\infty\,$, 
the magnetized Schwarzschild metric (\ref{3.10}) approaches the metric of the Melvin magnetic universe.
The energy within a sphere of radius $r_{0}$ for Melvin's universe 
can be easily obtained from Eq. (\ref{4.7}) taking $m = 0\,$. We see that due to the 
logarithmic term in Eq. (\ref{4.7}) it typically reflects the axial symmetry of the spacetime, which 
corresponds to the asymptotic direction of the cosmic magnetic field.    

\section{Final Remarks}

\noindent

We have studied the distribution of gravitational energy in the spacetime 
of a magnetized Schwarzschild black hole. For this we have applied the 
definition for gravitational energy-momentum established in the context of TEGR as 
a direct consequence of the field equation of the theory, which turns out 
to be equivalent to Einstein's field equation of General Relativity. We have computed 
the energy enclosed by a spherical surface involving the black hole. 
In particular, we have computed the energy both in the weak-field limit (in which case 
the spacetime is nearly `Newtonian') and in the strong field regime (energy enclosed 
by the event horizon of the black hole). The results clearly show the gravitational 
effect of the magnetic field, {\it i.e.}, the contribution of the magnetic field to the 
gravitational energy. This complies with the principle that any form of energy should 
contribute to gravity. In this sense, the $a=(0)$ component in the definition given by Eq. 
(\ref{def-energia-mom}) can be interpreted in the present case as giving the effective (or active) 
gravitational mass of the spacetime. 

The analysis of the gravitational energy enclosed by the 
event horizon of a magnetized Schwarzschild black hole has not been considered in the literature, as far as we know. 
Even in Ref. \cite{Dadhich}, in which the gravitational energy for a closed 2-surface in the spacetime of a magnetized 
Schwarzschild black hole is computed, by a means of a modification of the Komar charge, the energy enclosed by the event horizon 
is not considered, although it can readly be evaluted from Eq. (4.12) of that Ref. \cite{Dadhich}. 
It turns out that, according to the latter equation of Ref. \cite{Dadhich}, the energy 
enclosed by the horizon is just $m$, what means that the magnetic field has the effect of decreasing the energy enclosed by the horizon, 
as compared to the case of absence of magnetic field (as is well-known, in the Schwarzschild vacuum case, the value for the energy 
enclosed by the horizon is $2m$, as given, for instance, by the Brown-York quasilocal energy, which, as we have seen below of Eq. (24), 
is the same value given by energy definition in the TEGR). 
Nonetheless, contrary to the result of Ref. \cite{Dadhich}, for a magnetized Schwarzschild black hole, we have just obtained an increasing 
in the energy, according to our Eq. (22). This increasing makes sense physically, as the magnetic field contributes to the gravitational 
field, as it distorts the spacetime together with the black hole mass. 

We have thus properly addressed the gravitational energy distribution in the spacetime of magnetized Schwarzschild black hole. In particular, 
the energy associated with its event horizon, which is important for itself and mainly in view of studying the thermodynamic behaviour of this 
kind of black hole, or even in view of astrophysical motivations. We finally remark that although we have considered here a stationary spacetime, 
the definition for gravitational energy that arises in the framework of the TEGR is generally valid. In fact, the latter has been applied to nonstationary spacetimes \cite{Maluf-Faria-CB2003},\cite{Maluf-Faria2004},\cite{Maluf-Ulhoa-PRD2009}. 

\bigskip

\textbf{ACKNOWLEDGEMENTS}

We would like to thank the two anonymous referees for their remarks and questions that led to 
improvements on some points in the presentation of the work.

\end{document}